\documentclass[twocolumn,aps,prd,preprintnumbers,showpacs,superscriptaddress,nofootinbib,amsmath,amssymb,floats,floatfix,showkeys,notitlepage,longbibliography]{revtex4-1}
\usepackage{graphicx}	
\usepackage{amssymb}
\usepackage{dcolumn}
\usepackage{mathtools}
\usepackage{amsmath}
\usepackage{xcolor}
\usepackage{color}
\usepackage{orcidlink}
\usepackage{theorem}
\usepackage{subfigure, rotating, bm, array}
\hypersetup{linkcolor=blue, citecolor=blue,urlcolor=blue} 

\usepackage[english]{babel}           

\begin{document}
\renewcommand{\refname}{References}

\title{
Energy Extraction and Evolution of Regular Black Holes: The Case of Bardeen Spacetime
}

\author{Vitalii Vertogradov}
\affiliation{Physics Department, Herzen State Pedagogical University of Russia, 48 Moika Emb., Saint Petersburg 191186, Russia}
\affiliation{Center for Theoretical Physics, Khazar University, 41 Mehseti Street, Baku, AZ-1096, Azerbaijan.}
\affiliation{SPB branch of SAO RAS, 65 Pulkovskoe Rd, Saint Petersburg
196140, Russia}

\email{vdvertogradov@gmail.com}
\author{Angel Rincon \orcidlink{0000-0001-8069-9162}}
\affiliation{Departamento de Física Aplicada, Universidad de Alicante, Campus de San Vicente del Raspeig, E-03690 Alicante, Spain}
\affiliation{Research Centre for Theoretical Physics and Astrophysics, Institute of Physics, Silesian University in Opava,
Bezručovo náměstí 13, CZ-74601 Opava, Czech Republic}
\email{angel.rincon@ua.es}

\date{\today}

\begin{abstract}
This paper examines regular black holes, in particular the Bardeen spacetime where singularities are replaced by non-singular cores. It explores the energy extraction through the charged Penrose process and shows that magnetic charge evaporation can drive a regular black hole towards a singularity. Two evaporation models are proposed, dealing with charge loss and combined charge evaporation with mass accretion, providing insights into the evolution and stability of regular black holes.

\end{abstract}

\maketitle

\section{Introduction}

Recent experimental observations of black hole shadows in the centres of the M87~\cite{bib:eht1, bib:eht2} and our Milky Way ~\cite{bib:eht3} granted black holes the status of real astrophysical objects. However, classical solutions to the Einstein equations, under generic conditions regarding the matter content, possess past and future singularities, which are usually hidden by the horizon.
Furthermore, the famous singularity theorem~\cite{bib:singularity} states that the gravitational collapse should lead, under certain conditions, to the singularity formation. According to the cosmic censorship conjecture, the singularity must be hidden by the event horizon. 
Nevertheless, recent developments in the gravitational collapse revealed a huge number of models, in which, under physically-relevant conditions, this process might lead to the naked singularity formation~\cite{bib:joshi, bib:joshi_review, bib:zek, bib:zek2, bib:maharaj, bib:ver_eternal, bib:structure, bib:dis, bib:radial, bib:burst, bib:vertogradov2024pocs, bib:vertogradov2024grg}. 
The existence of a singularity underlines the inadequacy of general relativity to describe spacetime below some length scale. This fact motivated to seek non-singular (also called regular) solution of general relativity. 
In light of the fact that spacetime singularities reveal the limitations of general relativity, the inclusion of a novel ingredient is mandatory to cure such a problem. In such a sense, a quantum theory of gravity is a natural way of dealing with some of these weaknesses.
In the absence of a fully developed and reliable quantum gravity framework, a conventional approach is to consider models of singularity-free black holes (usually called ``regular" black holes).

It is worth noting here that regular black holes can be simply described as solutions of Einstein's field equations in which the central singularity is replaced by a non-singular core. 
This modification effectively circumvents the issue of infinite curvature (see, for example, \cite{Lan:2023cvz} and references therein).
It should be noted that the singularity involves two complementary aspects: the first concerns the divergence of the curvature invariants, while the second corresponds to the incompleteness of the geodesics. Therefore, to ensure that a black hole is indeed regular, it is more appropriate to combine two tests.
First, it should be established whether the curvature invariants are uniformly restricted to a specific universal value, as required by Markov's limiting curvature conjecture \cite{1982ZhPmR..36..214M,Frolov:1988vj,Frolov:2016pav,Chamseddine:2016ktu}). 
Second, the alternative strategy to determine a regular spacetime is based on geodesic completeness. The latter means that spacetime is regular if its null and time-like geodesics are complete, i.e., the affine parameter of a test particle does not terminate at a finite value (see \cite{Hawking:1973uf,Wald:1984rg} and references therein).
Thus, both criteria should be satisfied to obtain a regular black hole.

There are plenty of papers which contributed to the understanding of the so-called ``regular" black holes. For instance, Sakharov~\cite{bib:sakharov} and Gliner~\cite{bib:gliner} assumed that the singularities could be avoided if the vacuum is replaced by the vacuum medium described with de Sitter spacetime. The first non-singular model of a black hole was introduced by Bardeen~\cite{bib:bardeen}. The core of this solution is de Sitter, and this solution behaves like Schwarzschild black hole at infinity. The first interpretation of the matter content was given three decades after Bardeen's work. 
In addition, several works focus on the Bardeen black hole and account for progress using such a background. For instance in \cite{Kumar:2020uyz},  exact Bardeen black holes to the regularized Einstein–Gauss–Bonnet gravity minimally coupled with the nonlinear electrodynamics are obtained. Similarly, in \cite{Stuchlik:2014qja}, the circular geodesic motion of test particles and photons in the Bardeen and Ayon–Beato–Garcia geometry describing spherically symmetric regular black-hole or no-horizon spacetimes was properly investigated. Further interesting works that could help to clarify the importance of the Bardeen black hole can be found in references \cite{Kumar:2018vsm,Ghosh:2015pra}.
Ayon-Beato and García showed that this solution is supported by a magnetic monopole in the context of nonlinear electrodynamics~\cite{bib:garsiya}. This magnetic charge $g$ prevents the formation of a singularity inside a black hole.
It is undeniable that another work by Hayward \cite{Hayward:2005gi} had a remarkable impact (the so-called Hayward black hole). 
These regular metrics often involve modifications to Einstein's field equations to ensure that curvature invariants such as the Kretschmann and Ricci scalars remain finite at the core of the black hole.
Regular black holes have been investigated in detail from several different perspectives, ranging from non-linear electrodynamics \cite{Bronnikov:2000vy,Bronnikov:2000yz,Bronnikov:2017sgg,Contreras:2017eza,Panotopoulos:2019qjk}, alternative theories of gravity \cite{Eichhorn:2022bgu,Rodrigues:2018bdc,Misyura:2024fho,Olmo:2022cui,Bonanno:2023rzk}, semiclassical theory \cite{Spallucci:2017aod}, among other applications 
\cite{Balart:2024rtj,Ovgun:2024zmt,Balart:2023odm,Rincon:2020cos,Javed:2022rrs,Fan:2016hvf,bib:khlopov,bib:khlopov2,bib:khlopov3,bib:khlopov4,bib:khlopov5}. 
Recently, a new black hole with a de Sitter core was obtained ~\cite{bib:ali2024cqg} with Hagedorn fluid as a source. This black hole can describe both a singular and non-singular black holes, which can be used to distinguish them through shadow properties.

Black holes can be used as super colliders~\cite{bib:grib2020pocs}. Penrose showed that in the ergoregion of a rotating black hole might be particles with negative energy states, which allow the extraction of rotational energy from a black hole~\cite{bib:pen}, processes which reduce the angular momentum $a$ of a black hole. The detailed investigation of this process can be found in~\cite{bib:vertogradov2014mpla, bib:vertogradov2015gc, bib:zaslav_dirty, bib:vertogradov2024negative}. The analogous effect exists for charged Reissner-Nordström spacetime for charged particles~\cite{bib:rufini}. However, if in Kerr spacetime the ergoregion is a manifestation of a geometry, in the Reissner-Nordström black hole, the generalized ergoregion depends on the particle's properties. This process leads to the decrease of an electrical charge $q$ of a black hole~\cite{bib:zaslav_negative}. Hawking~\cite{bib:hawking} showed that a black hole can lose its mass through the Hawking radiation process. Motivated by these facts, we decided to consider the possibility of a process, analogous to the charged Penrose process, that would lead to the decreasing magnetic charge $g$ of a black hole.

This process can lead to growing curvature invariants such as the Ricci and Kretschmann scalars. The question is whether this process can lead to singularity formation, i.e., some transition from non-singular (i.e., regular) to singular Schwarzschild black hole. We construct several models of evaporating magnetic charge.
There are several compelling reasons to investigate Hawking evaporation in electrically and magnetically charged black holes. Notably, contrary to the earlier belief that magnetic charge stabilizes black holes, a classical instability in non-Abelian gauge theories enables magnetically charged black holes to fully evaporate via Hawking radiation, leaving behind non-singular magnetic monopoles, as highlighted in Ref.  \cite{1992GReGr..24.1203L}. This contrasts with electrically charged black holes, where charge neutralization facilitates complete evaporation, thus highlighting the pivotal role of magnetic monopole theories in black hole physics.
While these phenomena have been studied within the context of Maxwell electrodynamics, our paper focuses on nonlinear electrodynamics, a subject that deserves significant attention due to its unique theoretical implications.
The issue of instability in regular black holes has previously been addressed in works~\cite{bib:new1, bib:new2}. The formation of a singularity is associated with the instability in the inner horizon and mass inflation. 
The possibility that even quantum gravity may also fail to resolve the singularity was discussed in ~\cite{bib:new3}.
In this paper, we present a different mechanism that leads to the instability in the regular center by extracting energy via the magnetic Penrose process. This mechanism is applied to the Bardeen regular black hole; however, as our calculations show, it can be extended to any spherically symmetric regular black hole supported by nonlinear electrodynamics.

Last but not least, a magnetic black hole is a black hole with a magnetic charge, a solution of the Standard Model coupled to gravity. Unlike an electrically charged black hole, which can be neutralized in a conductive medium, a magnetically charged black hole remains stable in ordinary matter, which makes it even more intriguing.
Maldacena \cite{Maldacena:2020skw} notes that strong magnetic fields near the horizon can restore electroweak symmetry. For certain magnetic charges, the black hole carries a weak hypercharge magnetic field, surrounded by an electroweak corona where the Higgs field transitions from zero to its typical non-zero value.
Furthermore, magnetic black holes do not conserve baryon number, which could potentially catalyze proton decay. In this process, protons falling into the black hole could emerge as positrons.
As highlighted in \cite{Maldacena:2020skw}, fermions in these magnetic fields form Landau levels, with the lowest having zero energy and a degeneracy proportional to the electric charge. This results in light modes that can enter and exit the black hole, enhancing Hawking radiation by a factor of the black hole’s charge, thus accelerating its evaporation.
However, this paper omits this effect for future study, despite its significant importance.

This paper is organized as follows. In Sect.\eqref{Bardeen_sec} we briefly describe the main properties of a regular Bardeen spacetime. In Sect.\eqref{Trajec_sec} we consider the trajectories for charged particles with negative energy and prove the absence of closed orbits for such particles in the generalized ergoregion. In Sects \eqref{Penrose_sec} and \eqref{Energy_sec} we consider the Penrose process and calculate the efficiency of this process, respectively. Subsequently, In Sec. \eqref{Evaporation_sec} we consider different scenarios of the magnetic charge and mass evaporation. Finally, we discuss the results obtained in last section.

We will use the geometrized system of units throughout the paper. We also adopt the signature $\{-,+,+,+\}$.

\section{Bardeen spacetime} \label{Bardeen_sec}

By selecting the mass function to ensure the finiteness of pressure and energy density at the center, as well as the vanishing of the mass function at the origin, Bardeen found a solution that describes a regular black hole, now known as the Bardeen metric~\cite{bib:bardeen}. This metric has the following form:
\begin{equation} \label{eq:metric}
ds^2=-f(r)dt^2+f(r)^{-1}dr^2+r^2d\Omega^2 \,.
\end{equation}
Here $d\Omega^2=d\theta^2+\sin^2\theta d\varphi^2$ is the metric on the unit two-sphere. The lapse function $f(r)$ is given by:
\begin{equation} \label{eq:lapse}
f(r)=1-\frac{2Mr^2}{(r^2+g^2)^{\frac{3}{2}}} \,.
\end{equation}
Here, \( M \) represents the mass of the black hole, while the parameter \( g \) can be associated with the charge of a monopole in a nonlinear self-gravitating magnetic field. Due to the presence of this charge, it becomes possible to study effects analogous to those observed in charged black holes, such as the Reissner-Nordström spacetime. In particular, the \textbf{Banados-Silk-West effect} is also applicable in the case of the Bardeen black hole~\cite{bib:vertogradov2024gc}.

We aim to investigate the existence of electrically charged particles with negative energy in the field of a Bardeen black hole. To this end, we consider the motion of a charged particle with an electric charge \( q \). The spacetime described by \eqref{eq:metric} possesses two Killing vectors, \( \delta_t \) and \( \delta_\varphi \), associated with time translation and rotational symmetries, respectively. The presence of these two Killing vectors leads to two conserved quantities: the energy per unit mass \( E \) and the angular momentum per unit mass \( L \) of the particle.

The Lagrangian for an electrically charged particle moving in the equatorial plane \( \theta = \frac{\pi}{2} \), with the electromagnetic potential~\cite{bib:bardeen_test}:
\begin{equation} \label{eq:magnetic}
A_\alpha=\phi \delta^t_\alpha \,,
\end{equation}
is given by

\begin{equation} \label{eq:lag}
2\mathcal{L}=-f\left(\frac{dt}{d\lambda}\right)^2+f^{-1}\left(\frac{dr}{d\lambda}\right)^2+r^2\left(\frac{d\varphi}{d\lambda}\right)^2-2 q A_\alpha\frac{dx^\alpha}{d\lambda}.
\end{equation}
Here, \(\phi = \phi(g, r)\) is the scalar potential of the monopole field. Various expressions for this potential can be found in the literature; however, we will use a general form. This choice is also motivated by the fact that the Bardeen metric is used here merely as an example, and our subsequent reasoning can be applied to any regular black hole supported by nonlinear electrodynamics. The only requirement we impose is that the spacetime describing the black hole must be asymptotically flat, which places the following condition on the potential:
\begin{equation}
\lim\limits_{r\rightarrow 0} \phi(g, r) = 0.
\end{equation}
Additionally, since in most models of regular black holes supported by nonlinear electrodynamics, the magnetic monopole serves as the regularization parameter, the potential must monotonically decrease, i.e.,
\begin{equation}
\frac{\partial \phi}{\partial r} < 0.
\end{equation}

From the expression for the Lagrangian and the fact that the spacetime possesses time translation and spherical symmetries, we can define the energy \( E \) and angular momentum \( L \) of the charged particle as follows:
\begin{equation} \label{eq:energy}
E=f\frac{dt}{d\lambda}+q\phi\,,
\end{equation}
\begin{equation} \label{eq:angular}
L=r^2\frac{d\varphi}{d\lambda} \,.
\end{equation}
Here $\lambda$ is an affine parameter.

We are interested in timelike motion so the following condition $g_{ik}u^iu^k=-1$ must be held. By using \eqref{eq:energy} and \eqref{eq:angular} in this condition, one obtains the radial velocity component:
\begin{eqnarray} \label{eq:radial}
\left(\frac{dr}{d\lambda}\right)^2&=&V_{\text{eff}},\nonumber  \\
V_{\text{eff}}&=&\varepsilon^2-f\left(\frac{L^2}{r^2}+1\right),\nonumber \\
\varepsilon &=&E-q\phi\,.
\end{eqnarray}
We aim to determine whether it is possible for a charged particle to possess negative energy. To this end, we solve for the energy from equation \eqref{eq:radial}, which leads to the following expression:
\begin{equation} \label{eq:minenergy}
E=q\phi+ \sqrt{f\left(\frac{L^2}{r^2}+1\right)+\left(\frac{dr}{d\lambda}\right)^2}.
\end{equation}
When taking the square root, two signs generally appear in front of the root. However, we must discard the negative sign due to the requirement that the energy must be positive at infinity. 

\begin{figure}[t!]
\includegraphics[width=0.5\textwidth]{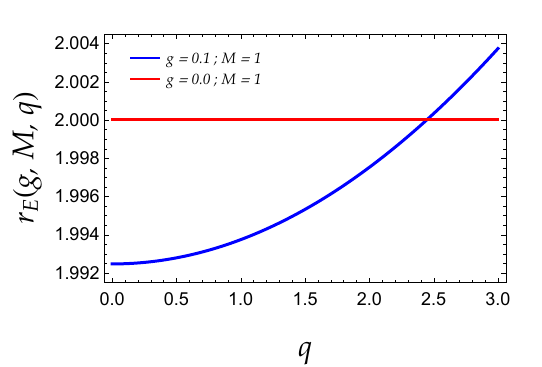}
\caption{Generalized ergosphere.}
\label{fig:1}
\end{figure}

From the expression for the energy \eqref{eq:minenergy}, it can be seen that the energy reaches its minimum values when the angular momentum is zero (\( L = 0 \)) and when the particle is at rest (\( \frac{dr}{d\lambda} = 0 \)). Taking this into account, we arrive at the following expression:
\begin{equation}
E=q\phi+\sqrt{f} \,.
\end{equation}
As can be easily seen from this expression, negative energies are possible only when \( q \) and \( \phi \) have opposite signs. Without loss of generality, we can assume that the electric charge is negative while the potential is positive. Furthermore, charged particles with negative energy cannot exist throughout the entire spacetime. The boundary of the region where they can exist is determined by the following formula:
\begin{equation} \label{eq:ergoregion}
-q\phi(g,r_E)= \sqrt{f(r_E)} \,.
\end{equation}
This border is so-called generalized ergosphere~\cite{bib:rufini}.  One should note that in comparison with the Kerr metric where the ergosphere is defined by geometrical properties only and its size doesn't depend on particle properties, in Bardeen spacetime the size of the generalized ergosphere depends on the electric charge of a particle. As a result every particle with negative energy state has their own region where it can exist. 
It is not possible to obtain a concrete expression for $r_E$ in closed form. However, \eqref{fig:1} shows how $r_E$ evolves with charge $q$ for two values of $g$: $g=0.0$ and $g=0.1$. The black hole mass is set to $M=1$.

\section{Trajectories for particles with negative energy} \label{Trajec_sec}

In this subsection, we investigate the trajectories of charged particles possessing negative energy. First, to perform the analysis, we must impose the forward-in-time condition:
\begin{equation}
\frac{dt}{d\lambda} > 0 \quad \Rightarrow \quad \varepsilon > 0.
\end{equation}
It is clear that a charged particle with negative energy cannot reach infinity, as this would imply \( \varepsilon < 0 \), violating the forward-in-time condition. Therefore, such a particle must have a turning point \( r = r_{tp} \) within the generalized ergoregion. Our goal is to prove that this turning point is unique within the generalized ergoregion. This, in turn, implies that the trajectories of such particles always fall into the event horizon, which is consistent with similar conclusions for particles with negative energy in Kerr~\cite{bib:vertogradov2014mpla, bib:vertogradov2015gc}, charged Vaidya~\cite{bib:ver_charged}, and Reissner-Nordström~\cite{bib:zaslav_negative} metrics.

The proof of the absence of a second turning point is straightforward, but some technical details are still required. A second turning point does not exist in the region \( r_{+} \leq r \leq r_{tp} \) if the quantity
\begin{equation} \label{eq:effectiveder}
\frac{V'_{\text{eff}}}{2}=b\frac{L^2}{r^3}+\varepsilon'\varepsilon-\frac{f'}{2} ,
\end{equation}
must be negative.  Where

\begin{equation} \label{eq:bvalue}
b=f-\frac{rf'}{2}  \,.
\end{equation}
$f'>0$ in the region $r_+\leq r \leq r_{tp}$, so one can conclude from \eqref{eq:bvalue} that
\begin{equation} \label{eq:inequality}
b\leq f \,.
\end{equation}
However, we do not know the sign of \( b \). If it is negative, the statement is proven, since \( \varepsilon' < 0 \). Moreover, the expression \eqref{eq:bvalue} is simply the condition for determining the radius of the photon sphere, and it is negative in the region between the photon sphere and the event horizon. This is because, although most black holes have more than one photon sphere, all but one—the one responsible for shadow formation—are hidden beneath the event horizon. Thus, if the turning point lies between the event horizon and the photon sphere, the trajectories of charged particles with negative energy always fall into the event horizon. However, as we have already mentioned, the existence of the generalized ergoregion depends on the parameters of the charged particle and is not manifestation of the spacetime itself, as is the case for rotating black holes. Therefore, we must consider positive values of the parameter \( b \).

To proceed with our analysis, we use the fact that the effective potential \( V_{\text{eff}}(r_{tp}) = 0 \) vanishes at the turning point. This leads us to the following inequality:
\begin{equation} \label{eq:help}
\frac{L^2}{r^2}\leq \frac{\varepsilon^2-f}{f} \,.
\end{equation}
Substituting \eqref{eq:help} into 
\eqref{eq:effectiveder},
one obtains the following restriction:
\begin{equation} \label{eq:result}
\frac{V'_{\text{eff}}}{2}\leq \frac{\alpha}{rf} \,.
\end{equation}
Here
\begin{equation}
\alpha=b\varepsilon^2-f\beta \,,
\end{equation}
and
\begin{equation} 
\beta=f-\varepsilon'\varepsilon \,.
\end{equation}
By using \eqref{eq:inequality} we can impose the following restriction on $\alpha$:
\begin{equation}
\alpha \leq f \gamma \,.
\end{equation}
Where
\begin{equation}
\gamma=\varepsilon^2+\varepsilon'\varepsilon r-f \,.
\end{equation}

Here, it is important to note that \( \varepsilon > 0 \) due to the forward-in-time condition, and \( \varepsilon' < 0 \) due to the constraints imposed on the potential \( \phi \). Consequently, the second term on the right-hand side is negative, and we need to analyze the expression \( \varepsilon^2 - f \).

We note that the boundary of the generalized ergoregion \eqref{eq:ergoregion} requires the equality \( -q\phi = \sqrt{f} \). However, since the particle's energy is negative (\( E < 0 \)), it follows that \( \varepsilon < \sqrt{f} \), which implies \( \varepsilon^2 - f \leq 0 \). This leads us to the conclusion that
\begin{equation}
\gamma<0 \implies \alpha<0 \implies \frac{V'_{\text{eff}}}{2}< 0  \,.
\end{equation}
This proves our assertion that the trajectory of a charged particle with negative energy has only one turning point within the generalized ergoregion. This, in turn, implies that all such trajectories inevitably fall into the black hole's event horizon.

We also note an important property: throughout our analysis, we have used the potential \( \phi \) and the lapse function \( f \) in their most general form. Consequently, our results are applicable to any asymptotically flat black hole supported by nonlinear electrodynamics.
\section{The Penrose process} \label{Penrose_sec}

In this section we investigate the energy extraction process which was proposed by Penrose~\cite{bib:pen}. We base our analysis on the following model: a negatively charged particle enters the generalized ergoregion, where, as a result of a collision or decay, it splits into two particles. One of these particles escapes to an observer located at \( R = r_0 \), carrying more energy than the original particle. The other particle falls into the event horizon, possessing negative energy. We denote the original particle with index \( 0 \), the particle with negative energy with index \( 1 \), and the escaping particle with index \( 2 \).

We again restrict our analysis to motion in the equatorial plane \( \theta = \frac{\pi}{2} \), which implies that the particle decays at the point \( \{R^*, \varphi^*\} \) located within the generalized ergoregion.

All parameters describing the particles—such as energy, charge, and momentum—obey conservation laws. Thus, the law of charge conservation states:

\begin{equation}
q_0 m_0=q_1 m_1+q_2 m_2 \,.
\end{equation}

The conservation of the four-momentum reads:
\begin{equation} \label{eq:cmomentum}
p^\mu_0=p^\mu_1+p^\mu_2 \,.
\end{equation}
The conservation of 4-momentum \eqref{eq:cmomentum} has direct physical implications. The temporal component of the equation represents the conservation of energy, the \( r \)-component represents the conservation of momentum, and the \( \varphi \)-component represents the conservation of angular momentum. All of these can be written as:
\begin{equation} \label{eq:cenergy}
E_0m_0=E_1m_1+E_2m_2 \,,
\end{equation}
\begin{equation} \label{eq:cmomenta}
m_0\left(\frac{dr}{d\lambda} \right)_0=m_1\left(\frac{dr}{d\lambda}\right)_1+m_2\left(\frac{dr}{d\lambda}\right)_2 \,,
\end{equation}
\begin{equation}
m_0L_0=m_1L_1+m_2L_2 \,.
\end{equation}

We note that all quantities are evaluated at the point of particle decay. To proceed with our analysis, we square the expression \eqref{eq:cmomentum} and, taking into account the fact that the 4-momentum is a timelike vector and that \( |p^\mu_i|^2 = m^2_i \), we arrive at the inequality:
\begin{equation} \label{eq:1mass}
m_0^2\geq m_1^2+m_2^2 \,.
\end{equation}
Now, using the expression \eqref{eq:cenergy}, we can estimate the energy carried to infinity by particle \( 2 \):
\begin{equation} \label{eq:process}
m_2E_2=m_0E_0-E_1m_1 \,.
\end{equation}
As we have previously established, charged particles with negative energy can exist within the generalized ergoregion. This implies that if, as we initially assumed, a particle with negative energy is produced at the decay point, then the formula \eqref{eq:process} describes an energy extraction process, which is a charged analog of the Penrose process. Furthermore, as noted in the previous section, particles with negative energy always fall into the black hole's event horizon. Consequently, the energy extraction occurs at the expense of reducing the black hole's charge.

We now consider the question of whether particle $2$ can indeed escape to infinity, which is necessary for energy extraction.
Without loss of generality, we assume that $q < 0$ and $\phi > 0$. To prove that particle $2$ escapes to infinity, we examine the radial geodesic equation:
\begin{equation}
\left(\frac{dr}{d\lambda}\right)^2 = \left(E - q\phi\right)^2 - f\left(\frac{L^2}{r^2} + 1\right) = -V_{\text{eff}},
\end{equation}
and show that the right-hand side remains positive and does not vanish on the interval $r \in [r_*, \infty)$, where $r_*$ corresponds to the point at which particle $0$ decays. Here, $V_{\text{eff}}$ denotes the effective potential.

For simplicity, we may set $L \equiv 0$. Clearly, $V_{\text{eff}}(r_*) > 0$, and as $r \to \infty$, with $E = 1$, the effective potential vanishes.

To demonstrate that the effective potential $V_{\text{eff}}$ is a monotonically decreasing function on the interval $r \in [r_*, \infty)$, or equivalently, that the particle has no turning points, it suffices to show that the derivative of the effective potential does not vanish on this interval.

Thus, we compute:
\begin{equation}
V'_{\text{eff}} = -2q\phi'\left(E - q\phi\right) - f'.
\end{equation}

Since the potential must vanish at infinity, we have $\phi' < 0$, implying that the first term is always negative. Therefore, we need to show that $f' > 0$ on $r \in [r_*, \infty)$.
We compute:
\begin{equation}
f' = \frac{6Mr^3(r^2 - 2g^2)}{(r^2 + g^2)^{5/2}}.
\end{equation}

We observe that $f' = 0$ when $r = \sqrt{2}g$. Since the decay occurs outside the event horizon, we must verify that $r = \sqrt{2}g$ lies inside the event horizon.
Indeed, consider the value:
\begin{equation}
f(\sqrt{2}g) = 1 - \frac{4M}{\sqrt{27}g}.
\end{equation}

However, for a black hole to exist — i.e., for there to be two horizons (or one in the extremal case) — the parameters $g$ and $M$ must satisfy the condition:
\begin{equation}
g \leq \frac{4M}{\sqrt{27}},
\end{equation}
with equality holding only in the extremal case. Thus, for non-extremal black holes, $f' = 0$ occurs only inside the event horizon, and for extremal black holes, it vanishes exactly at the horizon.
Therefore, outside the event horizon, we have $V'_{\text{eff}} < 0$, meaning no turning points exist. Consequently, the particle can escape to infinity carrying away more energy than initial particle $0$  had.

\section{The efficiency of the process} \label{Energy_sec}
We have established that in regular black holes supported by nonlinear electrodynamics, charged particles with negative energy can exist. This fact implies the possibility of energy extraction from the black hole through a charged analog of the Penrose process. In this section, we will examine the efficiency of this process.

The efficiency $\eta$ of the Penrose process can be defined as:
\begin{equation} \label{eq:etta}
\eta \equiv \frac{m_2E_2}{m_0E_0}-1=-\frac{m_1E_1}{m_0E_0} \,.
\end{equation}
To maximize energy extraction, it is necessary to obtain the largest possible value of $E_2$ from the smallest possible initial energy $E_0$. We note that the minimum initial energy $E_0$ is achieved when the particle falls from infinity, where it was initially at rest. In geometrized units, this condition results in $E_0 = 1$. Additionally, we assume that particle $0$, upon entering the region of the generalized ergosphere, has zero angular momentum, i.e., $L_0 = 0$. From equation \eqref{eq:minenergy}, it is clear that the minimum negative energy is attained when the particle is at rest, i.e.,
\begin{equation} \label{eq:energy3}
\begin{split}
\left(\frac{dr}{d\lambda}\right)_1=\left(\frac{d\varphi}{d\lambda}\right)_1=0 \,, \\
E=q\phi+\sqrt{1-\frac{2Mr^2}{(r^2+g^2)^\frac{3}{2}}}.
\end{split}
\end{equation}

From the condition \eqref{eq:energy3} one can easily see that $L_0=L_1=L_2=0$. Now we should evaluate the mass $m_1$, for this purpose, let's express $m_2$ from \eqref{eq:cmomenta}, keeping in the mind that $\left( \frac{dr}{d\lambda}\right) _1=0$:
\begin{equation} \label{eq:mass2}
m_2^2=\left[ \frac{m_0\left( \frac{dr}{d\lambda}\right)_0}{\left(\frac{dr}{d\lambda}\right)_2}\right]^2 \,.
\end{equation}
To derive the constraints on the efficiency of the process, we will use equations \eqref{eq:1mass} and \eqref{eq:mass2}, which yield:
\begin{equation} \label{eq:maxenergy}
\eta < - \left[q\phi+\sqrt{ 1-\frac{2Mr^2}{(r^2+g^2)^{\frac{3}{2}}}} \right]\sqrt{1-\frac{\left(\frac{dr}{d\lambda}\right)_0^2}{\left (\frac{dr}{d\lambda} \right)_2^2}} .
\end{equation}
As can be seen from equation \eqref{eq:maxenergy}, the extracted energy carried away by particle $2$ cannot exceed the absolute value of the energy of the incoming particle $1$. Note that the efficiency of the process directly depends on the electric charge of the particle and the monopole potential. Additionally, it is important to highlight the obvious fact that if the radial components of the 4-velocity of the initial and outgoing particles are identical, no energy extraction occurs. Conversely, the greater the velocity of the outgoing particle compared to the initial one, the more energy it carries away. Therefore, for energy extraction to occur, the following condition must be satisfied:
\begin{equation}
\left|\left(\frac{dr}{d\lambda}\right)_2\right|>\left|\left(\frac{dr}{d\lambda}\right)_0\right| \,.
\end{equation}

Thus, the greater the velocity of the outgoing particle compared to the initial one, the more energy is extracted from the black hole. The maximum energy extraction occurs in the immediate vicinity of the event horizon, where
\begin{equation} \label{eq:1.1}
\eta < -q\phi(r_{max})
\sqrt{1-\frac{\left(\frac{dr}{d\lambda}\right)_0^2}{\left (\frac{dr}{d\lambda} \right)_2^2}} \,.
\end{equation}

\section{Model of evaporation} \label{Evaporation_sec}
In this section, we examine illustrative examples of charge evaporation. Two distinct scenarios will be considered: one in which the charge evaporates while the mass remains constant, and another in which the charge decreases while the mass increases. 

In our analysis, we assume a linear dependence of both charge and mass on the advanced time coordinate $v$. This particular functional form is primarily motivated by the fact that, under such a dependence, the Bardeen metric admits a conformal Killing vector field~\cite{bib:maharaj_conformal, bib:vertogradov2023mpla}. This property has significant advantages: it allows a transformation to conformally static coordinates, facilitating the computation of observable features of the black hole, such as its shadow. Moreover, it enables the diagonalization of the metric, which in turn permits the introduction of a physically measurable time coordinate $T$, replacing the Eddington advanced time $v$, which lacks direct physical measurability.

\subsection{The model of the magnetic charge evaporation}
We offer the following model of the magnetic charge evaporation. First of all, let's consider the Bardeen spacetime in the ingoing Eddington-Finkelstein coordinates $\{v\,, r\,, \theta\,, \varphi\}$
\begin{equation}
ds^2=-\left(1-\frac{2Mr^2}{(r^2+g^2)^{\frac{3}{2}}} \right) dv^2+2dvdr+r^2d\Omega^2 \,.
\end{equation}
Now let's consider the magnetic charge $g$ as a decreasing function of the advanced time $v$ i.e.
\begin{equation}
ds^2=-\left(1-\frac{2Mr^2}{(r^2+g^2(v))^{\frac{3}{2}}} \right) dv^2+2dvdr+r^2d\Omega^2 \,.
\end{equation}
The Kretschmann scalar for this metric in the limit $r\rightarrow 0$ is given by:
\begin{equation}
\lim\limits_{r\to 0}K=R^{iklm}R_{iklm}=\frac{96 M^2}{g^6(v)} \neq  \infty 
\end{equation}
if $g(v)\neq 0$. The energy-momentum tensor for this spacetime can be written as
\begin{equation}
T^{total}_{ik}=T^{Bardeen}_{ik}+T^{evaporation}_{ik}\,.
\end{equation}
Where $T^{Bardeen}_{ik}$ is the energy momentum tensor of the Bardeen spacetime with the only exception that $g\equiv g(v)$. This part doesn't contain the time derivatives. $T^{evaporation}_{ik}$ represents the energy-momentum tensor which corresponds to the magnetic charge evaporation. It has only one non-vanishing term

\begin{equation}
T^r_v=+\frac{1}{4\pi}\frac{3Mrg(v)\dot{g}(v)}{(r^2+g^2(v))^\frac{5}{2}} .
\end{equation}
One should note that $T^r_v\geq 0$ in order to satisfy the weak energy condition ($\dot{g}(v)<0$).

For the simple model of magnetic charge evaporation, we offer the following model:
\begin{itemize}
\item $g=g_0=\text{const}.$. for the time $-\infty<v<0$;
\item $g=g_0-\lambda v$,  ~~~~  $ \lambda>0$  for $0\leq v\leq \frac{g_0}{\lambda}$;
\item $g\equiv 0$ for $v\geq \frac{g_0}{\lambda}$.
\end{itemize}
During the evaporation process the $T^r_v$ component of the energy momentum-tensor has the following form:
\begin{equation}
T^r_v=\frac{1}{4\pi}\frac{3\lambda 
M
r(1-\lambda v)}{(r^2+(1-\lambda v)^2)^{\frac{5}{2}}} \,.
\end{equation}

\subsection{The model II: the accreting mass and charge evaporation}

The first model is idealized one because infalling charged particle should also increase the black hole mass. We can also consider the time-depended mas function $M=M(v)$ in order to construct the model of accreting $M$ and evaporating $g$.
\begin{equation}
ds^2=-\left(1-\frac{2M(v)r^2}{(r^2+g(v)^2)^{\frac{3}{2}}} \right) dv^2+2dvdr+r^2d\Omega^2 \,.
\end{equation}
Like in previous case, one has only one difference with usual Bardeen energy-momentum tensor i.e. the off-diagonal term:
\begin{align} \label{eq:wek1}
\begin{split}
T^r_v=\frac{1}{4\pi} &\frac{r}{(r^2+g(v)^2)^{\frac{5}{2}}} \times
\\
&\left [ \dot{M}(v)(r^2+g(v)^2)-3M(v)g(v)\dot{g}(v)\right] \,.
\end{split}
\end{align}
As one can see, we can't satisfy the weak energy condition for arbitrary choice of the mass $M(v)$ and charge $g(v)$ function. In general, one has the region where WEC is violated i.e.
\begin{equation} \label{eq:wec}
r<\sqrt{g\left( \frac{3M\dot{g}}{\dot{M}}-g\right)} \,.
\end{equation}
Here, for charge evaporation we can consider the same model like we did in the previous case.
\begin{itemize}
\item Initially one has $M(v)=M_0$ and $g(v)=g_0$;
\item during charged Penrose process we adopt the following mass function:
\begin{eqnarray}
g(v)&=&g_0-\lambda v,~~ \lambda>0,\nonumber \\
M(v)&=&M_0+\mu v,~~ \mu>0.
\end{eqnarray}
This process lasts at the period of time $0, \frac{g_0}{\lambda}$.
\item At the time $v=\frac{g_0}{\lambda}$ the charge is equal to zero and the mass becomes $M=M_0+\frac{\mu g_0}{\lambda}$ and resulting spacetime becomes Schwarzschild one.
\end{itemize}

A few words should be said regarding the choice of functions $M(v)$ and $g(v)$. The linear dependence is not accidental. When considering the formation and evaporation of a Bardeen black hole, a linear dependence leads to the fact that the Bardeen metric admits the existence of a conformal Killing vector field $K = v\frac{\partial}{\partial v} + r\frac{\partial}{\partial r}$~\cite{bib:maharaj_conformal, bib:vertogradov2023mpla}, which significantly simplifies the analysis and even allows for the calculation of the shadow of a dynamical Bardeen black hole in this case.

\begin{figure}[t!]
\includegraphics[width=0.5\textwidth]{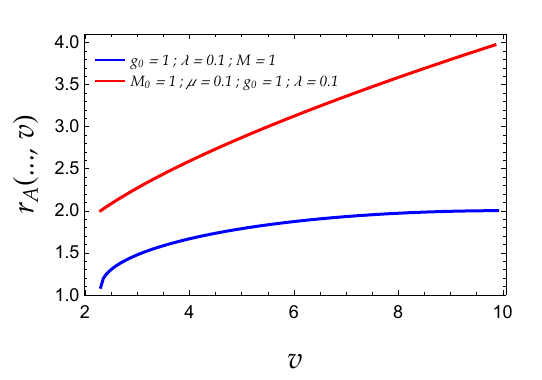}
\caption{Apparent horizon for different models of evaporation.}
\label{fig:2}
\end{figure}

In Fig.\eqref{fig:2}, we show the apparent horizon for the two evaporation models. The blue colour represents case A with $g=g_0 - \lambda v$, while the red colour represents case B with $g=g_0 - \lambda v$ and $M=M_0 + \mu v$. The values used appear in the figure.

\section{Discussions and concluding remarks}

In the present paper we have investigated the charged Penrose process for electrically charged test particles with magnetic charge.
We found that a charged particle with negative energy will fall onto a black hole. This particle must be charged oppositely to a black hole. From this fact, we can conclude that the extraction energy decreases the energy associated with the black hole, magnetic charge. 

\begin{figure}[t!]
\includegraphics[width=0.5\textwidth]{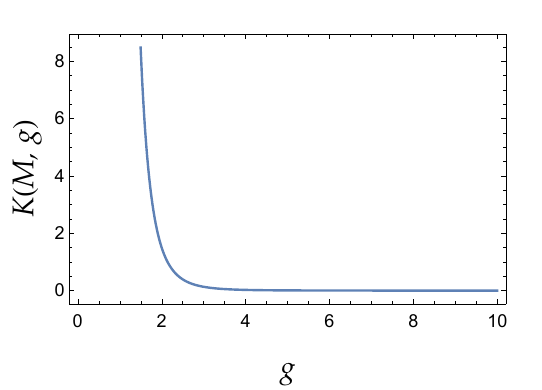}
\caption{Kretchmann scalar against $g$ for $M=1$.}
\label{fig:3}
\end{figure}

If the magnetic black hole charge is decreasing, then the curvature in the centre is increasing. If one considers the Kretschmann scalar 
\begin{equation}
K\equiv R_{iklm}R^{iklm} \,,
\end{equation}
in the centre 
\begin{equation}
\lim\limits_{r\rightarrow 0}K=\frac{96M^2}{g^6} \,,
\end{equation}
Then it becomes large as the magnetic charge becomes smaller. If the charge is evaporated by this process, the Kretschmann scalar will become divergent and this will lead to the formation of the singularity, 
as appears in Fig.\eqref{fig:3}
The question of singularity formation will be left aside for the time being, as a thorough investigation of the topology is required. There is also the very important question of whether the magnetic charge can evolve for a finite time. 

One important observation regarding black hole evaporation should be emphasized. In our analysis, we have considered the process under the assumption of a stellar-mass black hole. In this case, the generalized ergoregion is sufficiently large to allow the existence of charged particles with negative energy. However, when considering the evaporation of both charge and mass, the evaporation rates of these quantities must be taken into account, as derived in~\cite{bib:vertogradov2025plb}. If microscopic black holes could form during such a process, then at the Planck scale the size of the generalized ergoregion would become infinitesimally small. This would prevent the existence of charged particles with negative energy, rendering the results of this paper inapplicable to microscopic black holes.

Another important aspect is the fact that the Penrose magnetic effect requires the existence of hypothetical particles—magnetic monopoles. Naturally, we have not yet observed such particles, but their existence is predicted by several theoretical frameworks. In particular, Grand Unified Theory allows for the presence of magnetic monopoles as topological defects~\cite{thooft1974}. Moreover, string theory also accommodates magnetic monopoles~\cite{polchinski1996}.
An additional important issue concerns the formation of the regular Bardeen black hole. The key point here is that during the collapse of ordinary matter—devoid of magnetic monopoles—at critical densities, phase transitions may occur. As a result, baryonic matter can transform into a different state that supports a regular core within a regular black hole. A more detailed description of the underlying mechanism and potential observational signatures of such a collapse can be found in~\cite{bib:new4}.

It is important to note that the existence of the charged Penrose process, combined with the physically reasonable assumption that black holes should be neutral, makes nonlinear electrodynamics a poorly applicable source for regular black holes. The key issue is that during gravitational collapse, a de Sitter core, sourced by nonlinear electrodynamics, may form. However, the existence of such regular black holes would not be eternal. Due to neutralization—either through matter accretion or the charged Penrose effect—a singularity would form once the charge regularizing the black hole evaporates. 

In a recent study~\cite{bib:vertogradov2025podu}, it was demonstrated that a regular core could form through a more natural process. The proposed model in the paper considers the gravitational collapse of dust and radiation. Under the assumption that matter undergoes a phase transition into radiation at critical densities, a regular black hole could emerge as a result of gravitational collapse. Furthermore, it can be shown that Bardeen, Hayward, and Dymnikova regular black holes could form through this process without invoking nonlinear electrodynamics.

\section*{Acknowledgments}

The authors are grateful to the reviewers for their contributions, which have significantly improved the quality of this paper.
A. R. acknowledges financial support from Conselleria d'Educació, Cultura, Universitats i Ocupació de la Generalitat Valenciana through Prometeo Project CIPROM/2022/13.
A. R. is funded by the María Zambrano contract ZAMBRANO 21-25 (Spain) (with funding from NextGenerationEU).
%


\end{document}